\documentclass[10pt]{article}
\usepackage{lmodern}
\usepackage[margin= 1.2 in]{geometry}
\usepackage[utf8x]{inputenc}
\usepackage{hyperref}
\usepackage{url}
\usepackage{stmaryrd}
\usepackage{textgreek}
\usepackage{listings}
\usepackage{relsize}

\title{\textsf{Universal Algebra in UniMath}}
\author{Gianluca Amato \\
\textsl{Università di Chieti--Pescara} \and
Marco Maggesi\\
\textsl{Università di Firenze} \and
Maurizio Parton \\
\textsl{Università di Chieti--Pescara} \and
Cosimo Perini Brogi \\
\textsl{Universit\`a di Genova}}
\date{}

\begin{document}

\maketitle

We present an ongoing effort to implement Universal Algebra in the UniMath system \cite{UniMath}.
Our aim is to develop a general framework for formalizing and studying Universal Algebra in a proof assistant. By constituting a formal system for isolating the invariants of the theory we are interested in -- that is, general algebraic structures modulo isomorphism -- Univalent Mathematics seems to provide a suitable environment to carry on our endeavour.

After introducing the classical definition of signature, we give the related formalization of the category of algebras using the notion of a displayed category~\cite{lmcs:5252} over the univalent category of sets; in this way, we easily obtain a proof that the corresponding total category of algebras is univalent.

Our formalization also includes the notion of equations and varieties associated with a signature; as for the category of algebras, we construct the category of varieties over a given signature, and prove in a modular way that it is univalent by using the formalism of displayed categories. 

Our code is publicly available from 
\url{https://github.com/amato-gianluca/UniMath}. The revision discussed in this paper is tagged as \href{https://github.com/amato-gianluca/UniMath/tree/hott-uf-2020}{\texttt{hott-uf-2020}}. 



\subsection*{Approach and methodology}
Dependent types make easy to define the basic notions of arity and signature:
\begin{verbatim}
Definition Arity: UU := nat.
Definition signature: UU := ∑ (names: hSet), names → Arity.
Definition names (σ: signature): hSet := pr1 σ.
Definition arity {σ: signature} (nm: names σ): Arity := pr2 σ nm.
\end{verbatim}
The definition of an algebra over a signature follows straight:
\begin{verbatim}
Definition dom {σ: signature} (support: UU) (nm: names σ): UU
  := Vector support (arity nm).
Definition cod {σ: signature} (support: UU) (nm: names σ): UU
  := support.
Definition algebra (σ: signature): UU
  := ∑ (support: hSet), ∏ (nm: names σ), dom support nm → cod support nm.
\end{verbatim}

Things become less immediate when we move to terms, partly because UniMath does not natively support general inductive types. We represent each term of a signature as a sequence of function symbols (\texttt{names} in the code).  This sequence is thought to be executed by a stack machine (\texttt{oplist2status} in the code): each symbol of arity $n$ pops $n$ elements from the stack and pushes a new element at the top. A term is denoted thus by a sequence of function symbols that a stack-like machine can execute without stack underflow, returning a stack with a single element.
\begin{verbatim}
Definition isaterm (l: list (names σ)) := oplist2status l = statusok 1.
Definition term (σ: signature) := ∑ s: list (names σ), isaterm s.
\end{verbatim}

As a matter of fact, we find standard categorical presentations, though perspicaciously elegant in their abstractness, lacking a certain suitability for computerized mathematics. By contrast, our choice has been motivated by the intent of making \emph{all operations on terms evaluable in practice by the} UniMath \emph{environment}. In order to obtain such a result, having for these items both a recursion and an induction principle evaluable as functional terms of the formal system is mandatory.



This is in line with the theorem proving technique called (small-scale) reflection \cite{harrison1995metatheory}: algebraic reasoning is translated into syntactic manipulation which produces an algorithm written in the language of the object theory and run in that very form. In this way, we can really think of our (univalent) proofs about algebraic objects as programs. 





We have pursued this task according to a general methodological approach usually called Poincaré principle \cite{barendregt2013foundations}, whose distinctive cypher is the differentiation \emph{inside the same formal environment} of trivial computations from logic. Having proof-terms that the computational machinery of UniMath practically evaluates as correctly typed functions seems to fit that philosophy of mechanized mathematics better than just giving a formal counterpart of traditional mathematical notions that the computer cannot handle feasibly.

\subsection*{Results}
As stated before, the UniMath system suffices to formalize in a natural way the basics of Universal Algebra from an univalent perspective. In particular, the implementation of displayed categories has turned out essential in order to give a neat proof that the categories we are considering -- that is, algebras and varieties over a signature -- are univalent.

Also, we can easily prove that the type of algebra morphisms from an algebra to the term-algebra on the same signature $\sigma$ is contractible, so that the latter is initial in the category of algebras over $\sigma$.

On the side of the ``practicability'' of evaluation, we have been able to provide both a recursion and an induction principle on terms of a signature that \emph{do evaluate} w.r.t.~the normalization mechanism of UniMath -- which 
is a proper subsystem of the Coq proof-assistant. We take this as an indication that our methodological choice is really feasible.

In particular, the inductive hypothesis takes the following form:
\begin{verbatim}
Definition term_ind_HP (P: term σ → UU) :=
    ∏ (nm: names σ)
       (v: Vector (term σ) (arity nm))
       (IH: ∏ (i:⟦ arity nm ⟧), P (el v i))
    , P (build_term nm v).
\end{verbatim}
\texttt{Vector} is the type of finite-length vectors. It holds that 
\verb+Vector A n+ $\equiv$ \verb+A ×... × A+ with $n$ copies of \verb+A+. Moreover $\texttt{el v i}$ is the $i$-th element of the vector \texttt{v}. Then, the induction principles is stated as
\begin{verbatim}
Definition term_ind (P: term σ → UU) (R: term_ind_HP P) (t: term σ) : P t.
\end{verbatim}
Proving \verb+term_ind+ requires a very long proof by induction on the length of the list encoding the term.
Basically, the behaviour of term induction is characterized by the proof-term
\begin{verbatim}
Lemma term_ind_step (P: term σ → UU) (R: term_ind_HP P) (nm: names σ)
                    (v: Vector (term σ) (arity nm)) 
   : term_ind P R (build_term nm v)
     = R nm v (λ i: ⟦ arity nm ⟧, term_ind P R (el v i)).
\end{verbatim}
Note that \verb+term_ind_step+ is a propositional equality and does not hold at the judgmental level.

Among the authors' desiderata that still require some work we mention the construction of the initial object in the category of varieties, and the proofs of the classical theorems of homomorphisms along with the related constructions of quotients, products, and subvarieties.

\subsection*{An example}
Let us show a short example defining a signature and using term induction to construct a function over terms.
First of all, we define the signature for natural numbers:
\begin{verbatim}
Definition nat_signature: signature := stnset 2,, 
   el (vcons 0  (* arity of the zero term symbol •0 *)
      (vcons 1  (* arity of the succ term symbol •1 *)
       vnil)).
\end{verbatim}
where \texttt{stnset 2} is an \texttt{hSet} with 2 elements, denoted by \texttt{•0} and \texttt{•1}.
For example, the term representing the number four may be written as:
\begin{verbatim}
Definition four: list (names nat_signature) := •1 :: •1 :: •1 :: •1 :: •0 :: nil.
Definition term_four: term nat_signature := four,, idpath _.
\end{verbatim}

Now, let us show an example of recursive computation on terms.
We use \verb|term_ind| to define an operation \texttt{depth} which returns the depth of a term:
\begin{verbatim}
Definition depth: term σ → nat
  := term_ind (λ _, nat) (λ (nm: names σ) (vterm: Vector (term σ) (arity nm))
              (levels: ∏ i : ⟦ arity nm ⟧, (λ _ : term σ, nat) (el vterm i)),
              1 + vector_foldr max 0 (mk_vector levels)).
\end{verbatim}
We want to prove that the depth of \verb+term_four+ is $5$.
Since \texttt{depth} evaluates to numerals on closed terms, the proof is a straightforward application of the identity path:
\begin{verbatim}
Goal depth term_four = 5.
Proof. apply idpath. Qed.
\end{verbatim}

\subsection*{Related work}

A classical work on implementing Universal Algebra in dependent type theory is the one by Capretta~\cite{capretta:1999}, where he systematically uses setoids in Coq to handle equality on structures.  Another attempt, still based on setoids, has been recently carried on in Agda~\cite{gunther-gadea-pagano:2018}.  We share the feeling that Univalent Foundations provide a more principled approach to the task.

Initial semantics furnishes elegant techniques for studying induction and recursion principles in a general setting encompassing applications in programming languages and logic. Assuming univalence, steady research activity produced over the time a number of contributions to the UniMath library, see e.g.~\cite{ahrens_et_al:LIPIcs:2018:9671,ahrens_et_al:LIPIcs:2019:10513,ahrens_et_al:LIPIcs:2018:8472,10.1007/978-3-319-42432-3_2}.

Lynge's~\cite{lynge:2017} -- still under development in \cite{lynge-spitters:2019} -- seems to settle in a framework that more closely compares with ours.
Despite both our formalization and Lynge's one assume Univalent Mathematics as formal environment, the study we are proposing here differs from his one by adopting a more foundational perspective.
This point of view materialises in our choice of UniMath over CoqHoTT, which is the system adopted by Lynge for his encoding.
Moreover, our focus makes the implementation we are proposing different also from categorical treatments mentioned above because of the care we have taken about making the constructions easily evaluable by the very normalization procedure of proof-terms. 





\nocite{capretta:1999,lynge:2017,lynge-spitters:2019,gunther-gadea-pagano:2018}

\bibliographystyle{plain}
\bibliography{biblio}

\begin{thebibliography}{10}

\bibitem{ahrens_et_al:LIPIcs:2018:9671}
Benedikt Ahrens, Andr{\'e} Hirschowitz, Ambroise Lafont, and Marco Maggesi.
\newblock {High-Level Signatures and Initial Semantics}.
\newblock In Dan Ghica and Achim Jung, editors, {\em 27th EACSL Annual
  Conference on Computer Science Logic (CSL 2018)}, volume 119 of {\em Leibniz
  International Proceedings in Informatics (LIPIcs)}, pages 4:1--4:22,
  Dagstuhl, Germany, 2018. Schloss Dagstuhl--Leibniz-Zentrum fuer Informatik.

\bibitem{ahrens_et_al:LIPIcs:2019:10513}
Benedikt Ahrens, Andr{\'e} Hirschowitz, Ambroise Lafont, and Marco Maggesi.
\newblock {Modular Specification of Monads Through Higher-Order Presentations}.
\newblock In Herman Geuvers, editor, {\em 4th International Conference on
  Formal Structures for Computation and Deduction (FSCD 2019)}, volume 131 of
  {\em Leibniz International Proceedings in Informatics (LIPIcs)}, pages
  6:1--6:19, Dagstuhl, Germany, 2019. Schloss Dagstuhl--Leibniz-Zentrum fuer
  Informatik.

\bibitem{lmcs:5252}
Benedikt Ahrens and Peter~LeFanu Lumsdaine.
\newblock {Displayed Categories}.
\newblock {\em {Logical Methods in Computer Science}}, {Volume 15, Issue 1},
  March 2019.

\bibitem{ahrens_et_al:LIPIcs:2018:8472}
Benedikt Ahrens and Ralph Matthes.
\newblock {Heterogeneous Substitution Systems Revisited}.
\newblock In Tarmo Uustalu, editor, {\em 21st International Conference on Types
  for Proofs and Programs (TYPES 2015)}, volume~69 of {\em Leibniz
  International Proceedings in Informatics (LIPIcs)}, pages 2:1--2:23,
  Dagstuhl, Germany, 2018. Schloss Dagstuhl--Leibniz-Zentrum fuer Informatik.

\bibitem{10.1007/978-3-319-42432-3_2}
Benedikt Ahrens and Anders M{\"o}rtberg.
\newblock Some wellfounded trees in unimath.
\newblock In Gert-Martin Greuel, Thorsten Koch, Peter Paule, and Andrew
  Sommese, editors, {\em Mathematical Software -- ICMS 2016}, pages 9--17,
  Cham, 2016. Springer International Publishing.

\bibitem{barendregt2013foundations}
Henk Barendregt.
\newblock Foundations of mathematics from the perspective of computer
  verification.
\newblock In {\em Mathematics, Computer Science and Logic-A Never Ending
  Story}, pages 1--49. Springer, 2013.

\bibitem{capretta:1999}
Venanzio Capretta.
\newblock Universal algebra in type theory.
\newblock In Yves Bertot, Gilles Dowek, Andr\'e Hirschowits, Christine Paulin,
  and Laurent Th\'ery, editors, {\em Theorem Proving in Higher Order Logics,
  12th International Conference, TPHOLs '99}, volume 1690 of {\em LNCS}, pages
  131--148. Springer, 1999.

\bibitem{gunther-gadea-pagano:2018}
Emmanuel Gunther, Alejandro Gadea, and Miguel Pagano.
\newblock Formalization of universal algebra in {Agda}.
\newblock {\em Electronic Notes in Theoretical Computer Science}, 338:147--166,
  2018.

\bibitem{harrison1995metatheory}
John Harrison.
\newblock Metatheory and reflection in theorem proving: A survey and critique.
\newblock Technical Report CRC-053, SRI Cambridge, Millers Yard, Cambridge, UK,
  1995.
\newblock Available on the Web as
  \url{http://www.cl.cam.ac.uk/~jrh13/papers/reflect.pdf}.

\bibitem{lynge:2017}
Andreas Lynge.
\newblock Universal algebra in {HoTT}, 2017.
\newblock Bachelor's thesis, Department of Mathematics, Aarhus University.

\bibitem{lynge-spitters:2019}
Andreas Lynge and Bas Spitters.
\newblock Universal algebra in {HoTT}.
\newblock In {\em TYPES 2019, 25th International Conference on Types for Proofs
  and Programs}, 2019.

\bibitem{UniMath}
Vladimir Voevodsky, Benedikt Ahrens, Daniel Grayson, et~al.
\newblock {UniMath --- a computer-checked library of univalent mathematics}.
\newblock {Available} at \url{https://github.com/UniMath/UniMath}.

\end{thebibliography}

\end{document}